\documentclass[12pt]{article}
\usepackage{amssymb}
\usepackage{latexsym}
\newcommand{\BEQ}{\begin{equation}}
\newcommand{\EEQ}{\end{equation}}
\newcommand{\BEQA}{\begin{eqnarray}}
\newcommand{\EEQA}{\end{eqnarray}}
\newcommand{\BEQS}{\begin{displaymath}}
\newcommand{\EEQS}{\end{displaymath}}
\newcommand{\BEQAS}{\begin{eqnarray*}}
\newcommand{\EEQAS}{\end{eqnarray*}}
\newcommand{\BEI}{\begin{itemize}}
\newcommand{\EEI}{\end{itemize}}
\newcommand{\BED}{\begin{description}}
\newcommand{\EED}{\end{description}}
\newcommand{\BEN}{\begin{enumerate}}
\newcommand{\EEN}{\end{enumerate}}

\title{\Huge The Heat-Kernel in a Schwarzschild Geometry and the Casimir Energy}
\author{\large Frank Antonsen \\ University of Copenhagen \\ Niels Bohr 
Institute\\ Blegdamsvej 17, DK-2100 Copenhagen \O, Denmark
}
\date{}
\begin{document}
\maketitle

\begin{abstract}
We obtain an hybrid expression for the heat-kernel, and from that the 
density of the free energy, for
a minimally coupled scalar field in a Schwarzschild geometry at finite
temperature. This gives us the zero-point energy
density as a function of the distance from the massive object generating the
gravitational field. The contribution to the zero-point energy due to the 
curvature is extracted too, in this way arriving at
a renormalised expression for the energy density (the Casimir energy density). 
We use this to find an expression for other physical quantities: internal 
energy, pressure and entropy. It turns out that the disturbance of the 
surrounding vacuum generates entropy. For $\beta$ small the entropy is
positive for $r>2M$. We also find that the internal energy can be negative
outside the horizon pointing to the existence of bound states. The total
internal energy inside the horizon turns out to be finite but complex,
the imaginary part to be interpreted as responsible for particle
creation. 
\end{abstract}

\section{Introduction}
One of the most important quantum physical quantities is probably the 
zero-point energy. It is needed, for instance, when we want to study 
back-reaction, i.e., the influence the
matter fields moving in a curved back-ground assert on the back-ground 
geometry itself. This would be done by solving the Einstein equations with 
the expectation value of the energy-momentum tensor as source:
\BEQS
    G_{\mu\nu} = R_{\mu\nu}-\frac{1}{2}g_{\mu\nu}R = \langle T_{\mu\nu}\rangle,
\EEQS
where we have chosen units such that $\kappa=\hbar=c =1$. It is also important
for the study of renormalisation properties of a quantum field theory in a
curved space-time.\\
In order to find this quantity, we evaluate the heat-kernel corresponding 
to the
equations of motion for a scalar field. This calculation is carried out at
finite temperature. The integral of this heat-kernel with respect to some
fictitious fifth coordinate, $\sigma$, 
will give the density of Helmholtz' free energy (its
integral over the entire space-time will give the zeta-function, which is
essentially just Helmholtz' free energy). The resulting expression is 
regularised 
and renormalised in the subsequent sections.\\
We do this by using the spherical symmetry of the spacetime in order to 
collect all the unknown bits of the heat kernel into a function $g_{nl}(r,r';
\sigma)$ depending only upon the radial coordinates and $\sigma$. A recursive
relation for an asympototic expansion of this unknow function can be found and
solved, thereby allowing us to find the heat kernel.\\
From the freen energy one can derive expressions for the entropy and the
pressure by using standard thermodynamic relations. We find these and comment
on their meaning. We also show that, to lowest order in the mass generating
the Schwarzschild geometry, the free energy is that of an infinite family of
particles moving in one dimension with an $r$-dependent mass.

\section{Set-Up}
We consider a minimally coupled scalar field $\phi$ moving in a Schwarzschild
back-ground, hence the action is
\BEQ
    S=\frac{1}{2}\int(\partial_\mu \phi\partial_\nu\phi g^{\mu\nu} -\mu^2
    \phi^2)\sqrt{|g|}d^4x,
\EEQ
where the metric is given by the standard expression
\BEQS
    ds^2=\left(1-\frac{2M}{r}\right)dt^2-\left(1-\frac{2M}{r}\right)^{-1} 
	dr^2 - r^2d\theta^2-r^2\sin^2\theta d\phi^2,
\EEQS
where $M$ is the mass of the (classical) massive object generating the
Schwarzschild geometry, a black hole or a star, say. We will put the mass,
$\mu$, of the scalar field equal to zero. We can later reinsert it if we 
find it desirable.\\
The d'Alembertian becomes
\BEQS
    \Box = \frac{1}{h}\frac{\partial^2}{\partial
    t^2}-\frac{1}{r^2}\frac{\partial}{\partial r}r^2h\frac{\partial}{\partial r}
    -\frac{L^2}{r^2},
\EEQS
where $h(r)=(1-2M/r)$ and $L^2$ is the square of the angular momentum operator.
A finite temperature can be included by complexifying the time coordinate $t$,
$t\mapsto \tau$. As $\phi$ describes a Bose
field, it becomes periodic in $\tau$, $\phi(\tau+\beta)=\phi(\tau)$.
This implies that the time direction is not only complexified but also
compactified -- by appropriate scaling we can take $\tau$ to lie on the unit
circle $S^1$ (see e.g. Ramond \cite{Ram} or Itzykson and Zuber, \cite{IZ}, 
for further details on this). 
The action, furthermore, becomes ``euclideanised''.\\
Now, the partition function 
\BEQS
    Z=\int e^{-S} {\cal D}\phi = \int e^{-\int \frac{1}{2}\phi \Box \phi
    \sqrt{|g|}d^3xd\tau}{\cal D}\phi,
\EEQS
is simply a Gaussian and hence the functional integral can be carried out, the
result being
\BEQ
    Z = (\det\Box)^{-1/2}.
\EEQ
See for instance \cite{Ram}. The quantity we are particularly interested in, is
Helmholtz' free energy, $F$, which is defined as
\BEQ
    F=-\frac{1}{\beta}\ln Z =\frac{1}{2\beta}\ln\det\Box.
\EEQ
The internal energy and
pressure, which appear in the energy momentum tensor, are related to $F$ by the
usual thermodynamic relations, as is the entropy.

\section{Functional Determinants}
The major problem is apparently the calculation of $\det\Box$. For 
consistency, and in order to establish notation, 
I will give a short introduction to the topic here. Descriptions can be found 
in e.g. Hawking \cite{Haw} or Ramond \cite{Ram}.\\
A priori, the determinant of an operator $A$ must be given by the product 
of its eigenvalues $\lambda$
\BEQ
    \det A = \prod\lambda.
\EEQ
Now, obviously this is not an easy thing to calculate. This is where the zeta
function, $\zeta_A(s)$, and the heat-kernel, $G_A(x,x';\sigma)$, comes into
play. Define
\BEQ
    \zeta_A(s) = \sum \lambda^{-s},
\EEQ
then
\BEQ
    \det A = e^{-\zeta_A'(0)}.
\EEQ
The heat-kernel is defined through the differential equation
\BEQ
    A_xG_A(x,x';\sigma) = -\frac{\partial}{\partial\sigma}G_A(x,x';\sigma),
\EEQ
subject to the boundary condition
\BEQ
    \lim_{\sigma\rightarrow 0}G_A(x,x';\sigma) = \delta(x-x').
\EEQ
The reason for the terminology is transparent: when $A$ is 
$\frac{d^2}{d\theta^2}$,  the Laplacean on $S^1$, then $\zeta_A(s)\propto
\zeta(s)$, where $\zeta(s)=\sum_{n=1}^\infty n^{-s}$, is the Riemann
zeta-function, and when $A=\nabla^2$ and $\sigma$ is the temperature, then the
heat kernel satisfies the usual heat equation.\\
Denoting the eigenfunctions of $A$ by $\psi_\lambda$ we have
\BEQ
    G_A(x,x';\sigma) = \sum_\lambda \psi_\lambda(x)\psi_\lambda^*(x')
    e^{-\lambda\sigma},
\EEQ
and one easily proves the important relationship
\BEQ
    \zeta_A(s) = \frac{1}{\Gamma(s)}\int_0^\infty d\sigma \sigma^{s-1} \int d^4x
    G_A(x,x;\sigma).
\EEQ
Notice that the integral is only along the diagonal $x=x'$. We don't have to
know the eigenfunctions in order to solve the heat equation, in a
$d$-dimensional Euclidean space, for instance, one can show \cite{Ram}
\BEQS
    G_A(x,x';\sigma) = (4\pi\sigma)^{-d/2}e^{-\frac{(x-x')^2}{4\sigma}},
\EEQS
in Cartesian coordinates. We will later use a mixture of techniques to arrive
at a workable expression for the heat kernel. The angular and thermal part will
be handled by a mode sum, whereas we will have to make do with an asymptotic
expransion for the radial part in order to find the Casimir energy 
density.\\
We can arrive at an important interpretation of the value of the heat kernel
along the diagonal by noting
\BEQ
    f(x) \equiv-\frac{1}{2\beta}\left.\frac{\partial}{\partial s}\right|_{s=0}
    \frac{1}{\Gamma(s)}\int_0^\infty d\sigma \sigma^{s-1}G_A(x,x;\sigma),
\EEQ
is the density of Helmholtz' free energy
\BEQ
    F = \int f(x)\sqrt{|g|}d^4x,
\EEQ
and is thus exactly the quantity we are interested in calculating.

\section{Finding the Heat-Kernel}
We now return to our particular problem, the minimally coupled scalar 
field in a
Schwarzschild back-ground, i.e. the calculation of the determinant of the
d'Alembertian. For this particular operator we know that the eigenfunctions can
be written
\BEQ
    \psi_\lambda(\tau,r,\Omega) = e^{-i\omega_n\tau}Y_{lm}(\Omega) 
	g_\lambda(r), \label{eq:ef}
\EEQ
where $\Omega$ denotes the angles, and where $\omega_n$ is the {\em Matsubara
frequency}
\BEQ
    \omega_n = \frac{2\pi n}{\beta}\qquad n=0,\pm 1, \pm 2,... .
\EEQ
We have no intention of finding the eigenvalues $\lambda$, nor of finding
$g_\lambda(r)$, instead we will rewrite the d'Alembertian as
\BEQ
    \Box = -\frac{\omega_n^2}{h}-\frac{1}{r^2}\frac{\partial}{\partial r} r^2h
    \frac{\partial}{\partial r} - \frac{l(l+1)}{r^2}.
\EEQ
And the heat-kernel will be written as
\BEQ
    G(x,x';\sigma) = \sum_{nlm}g_{nl}(r,r';\sigma)Y_{lm}(\Omega) Y_{lm}^*
    (\Omega')e^{-\lambda_{nl}(r)\sigma-i\omega_n(\tau-\tau')},
	\label{eq:G}
\EEQ
where we have defined
\BEQ
    \lambda_{nl}(r) \equiv \frac{l(l+1)}{r^2}+\frac{\omega_n^2}{h}.
\EEQ
The reasoning behind this formula is as follows. The eigenfunctions can be
written according to (\ref{eq:ef}) as a product of a spherical harmonic, a
plane wave invloving the Matsubara frequency and the complexified time, and 
finally some unknown radial function. Clearly, we would expect the heat
kernel, being a sum of products of eigenfunctions, to have a similar form, but
since we cannot find the radial eigenvalues $\lambda$ we cannot simply use
(\ref{eq:ef}) to find $G$. On the other hand, it must be possible to expand it
on the functions $Y_{lm}(\Omega)Y_{lm}^*(\Omega')$ and $e^{-i\omega_n(\tau-
\tau')}$, hence the expression (\ref{eq:G}).\\
A mass, $\mu$, of the scalar field can be reinserted simply by adding 
$\mu^2$ to
$\lambda_{nl}(r)$. The heat equation determines the unknown functions
$g_{nl}$. We will Taylor expand these in $\sigma$, writing (asymptotic 
expansion)\footnote{For simplicity we will follow the common abuse of 
terminology and refer to this expression as ``the heat kernel'' although
it strictly speaking is only an asymptotic formula valid for $\sigma$ not
too large.}
\BEQ
    g_{nl}(r,r';\sigma) = \frac{1}{\sqrt{4\pi\sigma}}e^{-\frac{(r-r')^2}{4
    \sigma}}\sum_{k=0}^\infty a_k(r,r')\sigma^k,
\EEQ
where the first term is the form $g_{nl}$ would have in a flat space-time. The
boundary condition implies $a_0\equiv 1$. This asymptotic expansion is similar
to the standard Schwinger-DeWitt expansion \cite{BD,GMM}. The difference here
lies in the fact that we only make an asymptotic expansion of the radial part 
of the heat kernel, whereas we use a mode sum for the remaining part.
Our expansion is thus a hybrid, combining the asymptotic Schwinger-DeWitt
expansion with the exact mode sum.\\ 
Inserting this expansion into the heat
equation yields a recursion relation for the $a_k$'s. As we only need to know 
the
value of the heat-kernel along the diagonal $x=x'$, we will put $r=r'$ to arrive
at the following recursion relation by straightforward computation
\BEQ
    \left(1+k+\frac{M}{r}\right)a_{k+1} =- L_1 a_{k} -L_2 a_{k-1},
\EEQ
where we have defined the operators
\BEQA
    L_1 &=& \left(1-\frac{2M}{r}\right)\frac{d^2}{dr^2} + \left(\frac{2}{r} -
    \frac{8M}{r^2}\right)\frac{d}{dr}\nonumber\\
    &=& h\frac{d^2}{dr^2}+\frac{2}{r}(2h-1)\frac{d}{dr},\\
    L_2 &=& h\lambda_{nl}'\frac{d}{dr}+2h\lambda'_{nl}r^{-1} + h' \lambda_{nl}'
    + h\lambda_{nl}''.
\EEQA
Putting
\BEQ
    H = L_2 a_0 = 2h\lambda_{nl}'r^{-1}+h'\lambda_{nl}'+h\lambda_{nl}'',
\EEQ
the first few functions $a_k$ becomes
\BEQA
    a_0 &=& 1,\\
    a_1 &=& 0,\\
    a_2 &=& \frac{-2H}{5-h},\\
    a_3 &=& \frac{(-1)^22^2}{7-h}L_1\frac{H}{5-h},\\
    a_4 &=& \frac{(-1)^32^3}{9-h}L_1\frac{1}{7-h}L_1\frac{H}{5-h} +\frac{(-1)
    ^22^2}{9-h}L_2\frac{H}{5-h},\\
    a_5 &=& \frac{(-1)^42^4}{11-h}L_1\frac{1}{9-h}L_1\frac{1}{7-h}L_1\frac{H}
    {5-h} + \frac{(-1)^32^3}{11-h}L_1\frac{1}{9-h}L_2\frac{H}{5-h}+\nonumber\\
        &&\frac{(-1)^32^3}{11-h}L_2\frac{1}{7-h}L_1\frac{H}{5-h}.
\EEQA
There are singularities in these at $r=2M$, as one would expect. Noting that 
both $H$ and $L_2$ are linear in $\omega_n^2$, we see that $a_k$ can
contain all even powers of the Matsubara frequencies up to and including 
$\omega_n^{2[k/2]}$.\\
The method of summation of modes when calculating zeta functions have been 
applied in a number of other spacetimes by various authors, \cite{others},
and a number of explicit results exist for some rather simple cases. 

\section{The Free Energy Density}
We have now found an asymptotic expansion for the heat-kernel and are 
ready to integrate it. The $\sigma$
integration is trivial (it usually is), and will just give a sum of
$\Gamma$-functions. Explicitly
\BEQA
    \tilde{f}(x;s)&\equiv &\frac{1}{\Gamma(s)}\int_0^\infty d\sigma 
\sum_{nlm}|Y_{lm}(\Omega)|^2\frac{1}{\sqrt{4\pi}}\sum_{k=0}^\infty a_k
e^{-\lambda_{nl}(r)\sigma}\sigma^{s+k-3/2}\nonumber\\
    &=& \sum_{nlm}|Y_{lm}|^2\frac{1}{\sqrt{4\pi}}\sum_{k=0}^\infty a_k(r)
    \frac{\Gamma(s+k-\frac{1}{2})}{\Gamma(s)}\lambda_{nl}^{\frac{1}{2}-s-k}.
\EEQA
One should note that even though the asymptotic series is only valid for
$\sigma$ small, this integral still makes sense since the factor $\lambda_{nl}$
acts as a mass term regularisation. If we hadn't used our hybrid method we 
would have had to assume a mass for the scalar field in order to obtain a 
meaningfull integral. We can still do this, if we wish, by simply adding 
$\mu^2$
to $\lambda_{nl}$. The addition of such a finite mass, however, will not modify
the convergence properties of $\tilde{f}$, hence the succesive regularisations
to be performed below are still needed. This is so because only the sum over 
$n$ need to be regularised, and for fixed $l$, $\lambda_{nl}$ acts like a
finite mass irrespective of whether $\mu^2\neq 0$ or not.\\ 
For $k\neq 0$, differentiating this with respect to $s$ at $s=0$ simply 
amounts to removing the
$\Gamma$-function in the denominator and putting $s=0$ in the remaining terms.
The  $k=0$ term is singular and has to be treated separately. We have
\BEQS
    \tilde{f}_{k=0}(x;s) = \sum_{nlm} |Y_{lm}|^2\frac{1}{\sqrt{4\pi}}
    \frac{\Gamma(s-\frac{1}{2})}{\Gamma(s)}\lambda_{nl}^{\frac{1}{2}-s},
\EEQS
so we have a divergent sum (for $s=0$) in the numerator, namely $\sum_n 
\sqrt{z^2+4\pi^2 n^2}$, where $z^2=\frac{l(l+1)h(r)\beta^2}{r^2}$. We can 
carry out the sum for an arbitrary $s$, see e.g. Ramond \cite{Ram}:
\BEQAS
    \frac{1}{\Gamma(s)}\sum_{n=-\infty}^\infty \left(z^2+4\pi^2 n^2\right)^
      {\frac{1}{2}-s}
    &=& \frac{1}{\Gamma(s)}\left[ z^{1-2s}+\frac{2\zeta(2s+1)}{(4\pi^2)^
      {s-\frac{1}{2}}} +\right.\\
    &&\hspace{-20mm}\left. \frac{2}{\Gamma(s+\frac{1}{2})}\sum_{\nu=1}^\infty 
      \frac{x^{2\nu}}{\nu!}\frac{(-1)^\nu}
    {(4\pi^2)^{s+\nu-\frac{1}{2}}}\Gamma(s+\nu-\frac{1}{2})
      \zeta(2s+2\nu-1)\right].
\EEQAS
Differentiating and putting $s=0$ we obtain
\BEQS
    z+\frac{1}{2\pi^{3/2}}(\gamma-2\ln 2\pi)(4\pi-\frac{z^2}{\pi})
    +\frac{2}{\sqrt{\pi}}\sum_{\nu=2}^\infty\frac{z^{2\nu}}{\nu!}\frac{(-1)^\nu}
    {(4\pi^2)^{\nu-\frac{1}{2}}}\Gamma(\nu-\frac{1}{2})\zeta(2\nu-1),
\EEQS
which is finite. This expression is to be multiplied by $\beta^{-1}h^{-1/2}$.\\
The free energy density $f(x)$ is then
\BEQA
    f(x) &=& -\frac{1}{2\beta}\left.\frac{\partial\tilde{f}}
    {\partial s}\right|_{s=0}\nonumber\\
    &=& -\frac{1}{2\beta}\sum_{lm}|Y_{lm}|^2(4\pi)^{-1/2}\left[
      -2\sqrt{\pi}\sqrt{\frac{l(l+1)}{r^2}}-\right.\nonumber\\
    &&2\sqrt{\pi}\frac{1}{2\pi^{3/2}}(\gamma-2\ln 2\pi)\left(\frac{4\pi}
      {\sqrt{h}\beta}
    -\frac{l(l+1)\sqrt{h}\beta}{\pi r^2}\right)+\nonumber\\
    &&\left. 4\sqrt{\pi}\beta^{-1}h^{-1/2}\sum_{\nu=2}^\infty \frac{(-1)^\nu}
      {\nu!}\left(
      \frac{l(l+1)h\beta^2}{4\pi^2 r^2}\right)^\nu
    \Gamma(\nu-\frac{1}{2})\zeta(2\nu-1)\right]-\nonumber\\
    &&-\frac{1}{2\beta}\sum_{nlm}|Y_{lm}|^2(4\pi)^{-1/2}\sum_{k=2}^\infty 
    a_k(r)\lambda_{nl}(r)^{\frac{1}{2}-k}\Gamma(k-\frac{1}{2}).
\EEQA
On the face of it, this function depends on {\em all} the coordinates, but
looking closer we see that there is no explicit time-dependence, and using the
sum rule
\BEQS
    \sum_m Y_{lm}(\Omega)Y^*_{lm}(\Omega') = \frac{2l+1}{4\pi} P_l({\bf u} \cdot
    {\bf u}'),
\EEQS
where ${\bf u, u}'$ are unit vectors given by the solid angles $\Omega, 
\Omega'$, we see that \cite{GR}
\BEQAS
    \sum_m |Y_{lm}(\Omega)|^2 &=& \frac{2l+1}{4\pi}P_l(1)\\
    &=& \frac{2l+1}{2^{l+2}\pi}\sum_{\nu=0}^{\left[\frac{1}{2}l\right]} 
    \frac{(-1)^\nu (2l-2\nu)!}{\nu!(l-\nu)!(l-2\nu)!}.
\EEQAS
So the angular dependency also disappears. In accordance with what we 
would expect,
the energy density depends only on the radial coordinate. We will write the
density as a mode sum
\BEQ
    f(r) = \sum_l f_l(r),
\EEQ
where
\BEQA
    f_l(r) &\equiv& \left\{\pi^{-1}\beta^{-2}
    \frac{1}{2\beta}(4\pi)^{-3/2}(2l+1)2^{-l}\left(
    \sqrt{\frac{l(l+1)\beta^2}{r^2}}+\right.\right.\nonumber\\
    &&\frac{1}{2\pi^{3/2}}(\gamma-2\ln 2\pi)\left(\frac{4\pi}{\sqrt{h}\beta^2}
    -\frac{l(l+1)\sqrt{h}\beta^2}{\pi r^2}\right)+\nonumber\\
    &&\left. \frac{2}{\beta\sqrt{h\pi}}\sum_{\nu=2}^\infty \frac{(-1)^\nu}
      {\nu!}\left(\frac{l(l+1)h\beta^2}{4\pi^2 r^2}\right)^\nu
    \Gamma(\nu-\frac{1}{2})\zeta(2\nu-1)\right)-\nonumber\\
    &&\left.\frac{1}{2\beta}(4\pi)^{-3/2}(2l+1)2^{-l}\sum_{n=-\infty}^\infty
    \sum_{k=2}^\infty \Gamma(k-\frac{1}{2})a_k(r)\lambda_{nl}(r)^{1/2-k}
      \right\}\times\nonumber\\
    &&\sum_{\nu=0}^{[l/2]}\frac{(-1)^\nu(2l-2\nu)!}{\nu!(l-\nu)!(l-2\nu)!}.
\EEQA
The major problem here is that the coefficients $a_k(r)$ depend on $n$ through
the Matsubara frequencies. \\
This expression needs regularisation. For instance, when $l=0$ we encounter a 
singularity. Remembering that $a_k$ can include all
even powers of $n$ up to and including $n^{2[k/2]}$ we write
\BEQS
    a_k(r) = \sum_{t=0}^{\left[\frac{k}{2}\right]} b_k^{(t)}(r)n^{2t},
\EEQS
where $b_k^{(s)}$ is independent of $n$.
Defining $\lambda=4\pi^2/h\beta^2$ we then have
\BEQAS
    \sum_{n=-\infty}^\infty\sum_{k=2}^\infty \left.a_k\right|_{l=0}
    \Gamma(k-\frac{1}{2})
    \lambda_{n0}^{\frac{1}{2}-k} &=& 2\sum_{k=2}^\infty \Gamma(k-\frac{1}{2})
    \lambda^{\frac{1}{2}-k}\sum_{t=0}^{[k/2]}b_k^{(t)}(r)\sum_{n=1}^\infty 
      n^{2t+1-2k}\\
    &=& 2\sum_{k=2}^\infty \Gamma(k-\frac{1}{2})\lambda^{\frac{1}{2}-k} 
    \sum_{t=0}^{[k/2]}b_k^{(t)}(r)\zeta(2k-2t-1),
\EEQAS
which is divergent whenever $2k-2s-1=1$, as $s\leq \frac{1}{2}k$ this can
only happen when $k=2, s=1$.

\section{The Minkowski Space Contribution and the Casimir Energy Density}
To get a physical quantity we must subtract the Minkowski space contribution
\cite{BD,GMM}, because the physical energy must be such that it vanishes in
flat spacetime. Doing this we thereby get the ``Casimir energy density''. 
The relationship
between this Casimir energy density, $\zeta$-functions and cut-off 
regularisations has been studied in \cite{casimir}. This subtracting off the
flat spacetime contribution is sometimes enough to cancel divergences,
because any manifold looks locally like Minkowski spacetime and hence will
have the same leading divergence as in flat spacetime. It will turn out, 
however, that the Casimir energy thus obtained in this case is not finite
and still needs some regularisation. The subtraction off the flat contribution
should therefore not be seen as much as a regularisation/renormalisation as 
simply a normalisation, \cite{BD,GMM}.\\
The heat kernel for the d'Alembertian is already known in flat spacetime using
Cartesian coordinates, but this will not be useful to us here, since we need
to act on the asymptotic expansion (\ref{eq:G}). Hence what we need, is to find
a similar expression for the flat spacetime heat kernel using a mixture of mode
sum (angular coordinates and Matsubara frequencies) and asymptotic expansion
(radial coordinate).
We obtain this contribution
by letting $M\rightarrow 0$ in the metric (and hence the d'Alembertian 
etc.).\footnote{This is not the same as saying our result for the heat kernel
is smooth in $M$. We merely use the $M\rightarrow 0$ as a short hand for saying
that in that limit the metric tensor reduces to the Minkowski spacetime one -
with a compact time due to the finite temperature. The coefficients $a_k$ will
not a priori be smooth in this limit, however. Hence, new recursion relations
have to be found.}
Denoting the resulting coefficients by $\tilde{a}_k(r)$
we get the following much simpler recursion relation
\BEQA
    \tilde{a}_{k+1} &=& -\frac{1}{k+1}\tilde{L}_1\tilde{a}_k - \frac{1}{k+1}
    \tilde{L}_2\tilde{a}_{k-1}\nonumber\\
    &=& -\frac{1}{k+1}\left(\frac{d^2}{dr^2}+\frac{2}{r}\frac{d}{dr}\right) 
    \tilde{a}_k -\frac{2l(l+1)}{(k+1)r^3}\left(\frac{d}{dr}+\frac{2}{r} \right)
    \tilde{a}_{k-1}.
\EEQA
The solutions go like
\BEQ
    \tilde{a}_k(r) = \alpha_k(l)r^{-2k},
\EEQ
there is {\em no} dependency upon the frequency $\omega_n$. The coefficients
$\alpha_k$ satisfy (for $k\neq 0$)
\BEQ
    \alpha_{k+1} = -\frac{2k(2k-1)}{k+1}\alpha_k + \frac{4kl(l+1)}{k+1}
    \alpha_{k-1},
\EEQ
and $\alpha_0=1, \alpha_1=0$. Written down explicitly, the first few of the
$\tilde{a}_k$ are\footnote{If the mass, $\mu$, of the quanta of the scalar field
is non-zero, then these would be completely different. They would still be
independent of the Matsubara frequencies, though.}
\BEQAS
    \tilde{a}_0 &=& 1 = \alpha_0(l),\\
    \tilde{a}_1 &=& 0 = \alpha_1(l)r^{-2},\\
    \tilde{a}_2 &=& 2l(l+1)r^{-4} = \alpha_2(l)r^{-4},\\
    \tilde{a}_3 &=& -8l(l+1)r^{-6} = \alpha_3(l)r^{-6},\\
    \tilde{a}_4 &=& 12l(l+1)\left(5 -l(l+1)
    \right)r^{-8} = \alpha_4(l)r^{-8},\\
    \tilde{a}_5 &=& -672l(l+1)\left(1-\frac{17}{105} l(l+1)
    \right)r^{-10} = \alpha_5(l)r^{-10}.
\EEQAS
Denoting the corresponding free energy density by $f^{(0)}$, the Casimir
density is defined to be
\BEQ
    f^{\rm Cas}(r) = f(r)-f^{(0)}(r) = \sum_l f^{\rm Cas}_l(r).
\EEQ 
The very simple form the functions $\tilde{a}_k$ take on actually allow
us to carry out the summation over the Matsubara frequencies, using essentially
the same formulas as in the $M\neq 0$ case. Defining
\BEQ
    c_l \equiv -\frac{1}{2\beta}(4\pi)^{-3/2} (2l+1)2^{-l}
    \sum_{\nu=0}^{[l/2]}\frac{(-1)^\nu(2l-2\nu)!}{\nu!(l-\nu)!(l-2\nu)!},
\EEQ
the resulting free energy becomes
\BEQA
    f^{\rm Cas}_l &=& 2\sqrt{\pi}c_l\left[-
      \frac{\gamma-2\ln 2\pi}{2\pi^{5/2}}(z^2h^{-1/2}-\tilde{z}^2-
      \frac{4\pi^2}{\sqrt{h}\beta^2}+\frac{4\pi^2}{\beta^2})
    +\right.\nonumber\\
    && \frac{2}{\beta\sqrt{\pi}}\sum_{\nu=2}^\infty (z^{2\nu}h^{-1/2} -
      \tilde{z}^{2\nu})
    \frac{(-1)^\nu}{\nu ! (4\pi^2)^{\nu-\frac{1}{2}}} \Gamma(\nu-\frac{1}{2})
    \zeta(2\nu-1)+\sum_{k=2}^\infty\Gamma(k-\frac{1}{2})\times\nonumber\\
    &&\left(\sum_{n=-\infty}^\infty a_k(r)\lambda_{nl}^{\frac{1}{2}-k}-
    \alpha_k(l)r^{-2k}\beta^{2k-1}(2\pi)^{1-2k}\left\{\left(
    \frac{\tilde{z}}{2\pi}\right)^{1-2k} +
    2\zeta(2k-1)+\right.\right.\nonumber\\
    &&\left.\left.\left.\frac{2}{\Gamma(k-\frac{1}{2})}\sum_{\nu=1}^\infty
    \frac{\tilde{z}^{2\nu}}{\nu !}\frac{(-1)^\nu}{(2\pi)^{2\nu}}
    \Gamma(k+\nu-\frac{1}{2})\zeta(2k+2\nu-1)\right\}\right)\right],
\EEQA
where
\BEQ
    z^2 \equiv \frac{l(l+1)h(r)\beta^2}{r^2} \qquad\qquad
    \tilde{z}^2 \equiv \frac{l(l+1)\beta^2}{r^2},
\EEQ
and where we have used $zh^{-1/2}-\tilde{z}=0$. We should
note that $f_0^{(0)}$ contains a countable infinity of divergences, this 
time coming from the terms $\tilde{z}^{1-2k}\alpha_k(l)\sim l(l+1)^{1-2k+
[k/2]}$. We should furthermore notice that all the terms $z^{2\nu}h^{-1/2}-
\tilde{z}^{2\nu}$ are finite as $r\rightarrow 2M$, since $zh^{-1/2}\rightarrow 
0$ in this limit. The only divergent parts as $r\rightarrow 2M$ are the 
$a_k(r)\lambda^{\frac{1}{2}-k}$-terms. We can find an
improved expression for these by writing once more
\BEQS
    a_k(r)=\sum_{t=0}^{\left[\frac{k}{2}\right]}b_k^{(t)}(r)n^{2t},
\EEQS
we then have to perform sums of the form
\BEQ
    \xi_2(s,t;a;z) \equiv \sum_{n=-\infty}^\infty n^{2t}(z^2+an^2)^{-s},
\EEQ
with $a=2\pi$. For $t=0$ we know the result; it is just the formula we have 
been using a number of times by now. Denote this sum by $\xi_1(s;a;z)$. 
Explicitly it is
\BEQAS
    \xi_1(s;a;z) &\equiv& \sum_{n=-\infty}^\infty (z^2+an^2)^{-s}\\
    &=& z^{-2s}+\frac{2}{\Gamma(s)}
      \sum_{\nu=0}^\infty\frac{(-1)^\nu z^{2\nu}}{\nu! a^{s+\nu}}
      \Gamma(s+\nu)\zeta(2s+2\nu).
\EEQAS
For $t$ a positive integer (which is the only case we need to worry 
about) we notice the following relationship
\BEQ
    \xi_2(s,t;a;z) = \frac{(-1)^t\Gamma(s)}{\Gamma(s-t)
    }\frac{\partial^t}{\partial a^t}\xi_1(s-t;a;z).
\EEQ
Inserting the expression of $\xi_1$ we finally arrive at
\BEQ
    \xi_2(s,t;a;z) = \frac{2\Gamma(s)}{\Gamma(s-t)}\sum_{\nu=0}^\infty
    (-1)^\nu\frac{z^{2\nu}}{\nu!}\frac{\Gamma(s+\nu)}{\Gamma(s-t)}
    \zeta(2s-2t+2\nu)a^{-s-t-\nu}.
\EEQ
We have to evaluate this at $s=k-\frac{1}{2}, a=4\pi^2$ and multiply it by 
$(\beta^2 h)^{k-\frac{1}{2}}2\sqrt{\pi}c_l$ in order to get the contribution 
to the free  energy, which then reads
\BEQA
    f_l^{\rm Cas}(r) &=&2\sqrt{\pi}c_l\left[-\frac{\gamma-
      2\ln 2\pi}{2\pi^{5/2}}
    (z^2h^{-1/2}-\tilde{z}^2-\frac{4\pi^2}{\sqrt{h}\beta^2}+\frac{4\pi^2}
      {\beta^2})\right.+\nonumber\\
    &&\frac{2}{\beta\sqrt{\pi}}\sum_{\nu=2}^\infty \frac{(-1)^\nu}
      {\nu!(2\pi)^{2\nu-1}}
    (z^{2\nu}h^{-1/2}-\tilde{z}^{2\nu})\Gamma(\nu-\frac{1}{2})
      \zeta(2\nu-1)+\nonumber\\
    &&\sum_{k=2}^\infty\Gamma(k-\frac{1}{2})(2\pi)^{1-2k}\left(2\sum_{t=0}^{[k/2]}
    b_k^{(t)}(r)\frac{\Gamma(k-\frac{1}{2})}
      {\Gamma(k-\frac{1}{2}-t)}
    \sum_{\nu=0}^\infty(\beta^2h)^{k-\frac{1}{2}}(-1)^\nu\frac{z^{2\nu}}{\nu!}
      \times\right.\nonumber\\
    &&\frac{\Gamma(k-\frac{1}{2}+\nu)}{\Gamma(k-\frac{1}{2}-t)}
    (2\pi)^{-2k-2\nu+1-2t}\zeta(2\nu+2k-2t-1)-\nonumber\\
    &&\alpha_k(l)r^{-2k}\beta^{2k-1}(2\pi)^{1-2k}\left\{\left(
    \frac{\tilde{z}}{2\pi}\right)^{1-2k}+2\zeta(2k-1)
      +\right.\nonumber\\
    &&\left.\left.\left.\frac{2}{\Gamma(k-\frac{1}{2})}
      \sum_{\nu=1}^\infty
    \frac{(-1)^\nu\tilde{z}^{2\nu}}{\nu!(2\pi)^{2\nu}}
      \Gamma(k+\nu-\frac{1}{2})
    \zeta(2k+2\nu-1)\right\}\right)\right].
\EEQA
We note the presence of a singularity when $k=2, t=1$. It is the same kind of
singularity we encountered in the $l=0$ case. The regularisation of such
singularities is the subject of the next section.\\
It is an easy matter to derive a recursion relation for the new coefficients 
$b_k^{(t)}$; inserting $a_k=\sum_tb_k^{(t)}n^{2t}$ into the recursion 
relations for the $a_k$ yields
\BEQ
    \left(1+k+\frac{M}{r}\right)b_{k+1}^{(t)} = -L_1b_k^{(t)}-L_2^{(0)}
      b_{k-1}^{(t)}-L_2^{(1)}b_{k-1}^{(t-1)},
\EEQ
where we have written $L_2=L_2^{(0)}+n^2L_1^{(1)}$. Thus
\BEQAS
    L_2^{(0)} &=& -2\frac{l(l+1)}{r^3}\left(h\frac{d}{dr}+h'-h\frac{1}{r}
	\right),\\
    L_2^{(1)} &=& -\frac{4\pi^2}{h\beta^2}\left(h'\frac{d}{dr}+2h'\frac{1}{r}
    -\frac{h'{^2}}{h}+h''\right).
\EEQAS
The first few of these have been written out explicitly in the appendix.\\
Remembering that $b_k^{(t)}$ vanishes whenever $t<0$ or $t>\left[\frac{k}{2}
\right]$, we can calculate these coefficients systematically. We furthermore 
notice that $b_k^{(t)}$ is $\beta^{-2t}$ times some function which only 
depends upon the metric and its derivatives (i.e. on $h,h',h'',...$), we can 
thus write
\BEQ
    b_k^{(t)}(r,\beta) = \beta^{-2t}c_k^{(t)}(r),
\EEQ
in this way separating the coefficients into a purely thermal and a purely
geometric part.\\
Now, from Helmholtz' free energy density $f^{\rm Cas}$ we can calculate 
various interesting physical quantities, namely the renormalised expressions for
the (modes of) the pressure, $p_l$, internal energy density, $u_l$, and finally
the entropy density $s_l$ by
\BEQA
    p_l &=& -\left(\frac{\partial f^{\rm Cas}_l}{\partial V}\right)_\beta
    =-\frac{1}{4\pi r^2}\left(\frac{\partial f_l^{\rm Cas}(r)}{\partial r}
    \right)_\beta,\\
    u_l &=& \beta^2 f_l^{\rm Cas}+\beta s_l,\\
    s_l &=& \beta^2\left(\frac{\partial f_l^{\rm Cas}}{\partial\beta}
	\right)_r.
\EEQA
However, if one attempts to calculate, say, the entropy density, one will find
that it is divergent; thus there is still some regularisations to be done.

\section{The Final Regularisations}
As we saw, $f^{\rm Cas}$, had a singularity from $l=0$, stemming from 
two kinds of singularities in $f$ and $f^{(0)}$ which did not cancel each
other. We furthermore noticed a similar singularity in the $k=2,t=1$ term.
We will now return to this problem once more.\\
The singularities in the curved and flat space-time free energies
came from two different sources, one was $\zeta(1)$, i.e. the pole
of the Riemann $\zeta$-function, whereas the flat space-time
contribution had a countable infinity of poles $(l(l+1))^{-\nu}$.
We thus need two different ways of dealing with the problems.\\
It is well-known that singularities in the $\zeta$-function
regularization are to be removed by taking the appropriate
principal parts (see e.g. \cite{princip}), which for a meromorphic
function simply amounts to extracting the finite part near
a pole. Hence
\BEQA
    f_0(r) &=&-2\sqrt{\pi}c_0\left[\frac{1}{2}\sqrt{\pi}
    \left(\frac{4\pi^2}{h\beta^2}\right)^{-3/2}\left(
    \left.b_2^{(0)}\right|_{l=0}(r)\zeta(3)+\gamma 
    \left.b_2^{(1)}\right|_{l=0}(r)\right)+\right.\nonumber\\
    &&\left.\sum_{k=3}^\infty\Gamma(k-\frac{1}{2})
    \left(\frac{4\pi^2}{h\beta^2}\right)^{\frac{1}{2}-k}
    \sum_{t=0}^{\left[\frac{k}{2}\right]}\left.b_k^{(t)}\right|_{l=0}(r)
    \zeta(2k-2t-1)\right],
\EEQA
where we have written
\BEQ
    a_k(r) = \sum_{t=0}^{[k/2]}b_k^{(t)}(r)n^{2t},
\EEQ
as before. In obtaining this result we have used  \cite{GR,AS}
\BEQS
    \zeta(s)=\frac{1}{s-1}+\gamma+\sum_{n=1}^\infty (-1)^n
    \frac{\gamma_n}{n!}(s-1)^n.
\EEQS
Now
\BEQAS
    a_2 &=&\frac{-2H}{5-h}\\
    &=& \frac{8\pi^2 n^2}{(5-h)h\beta^2}\left( 2r^{-1}
    -h^{-1}(h')^2+h''\right)\mbox{   (for $l=0$)},
\EEQAS
so $\left.b_2^{(0)}\right|_{l=0}=0$\footnote{The $n=0$ contribution will always
be proportional to $l(l+1)$, as this is the only $\omega_n$-independent term on
the right hand side of the recursion relations. For $k=2$ we find explicitly
$b_2^{(0)} = 4r^{-4}l(l+1)(5-h)^{-1}(h'-h/r)$.}, leading to
\BEQA
    f_0(r) &=& \frac{1}{16\pi^{3/2}}\sqrt{h}\gamma (5-h)^{-1}
    (2r^{-1}-h^{-1}(h')^2+h'')+\nonumber\\
    && \hspace{-10mm}\frac{1}{8\beta\sqrt{\pi}}
    \sum_{k=3}^\infty\Gamma(k-\frac{1}{2})\left(
    \frac{\sqrt{h}\beta}{2\pi}\right)^{2k-1}\sum_{t=0}^{[k/2]}
    \left.b_k^{(t)}\right|_{l=0}(r)\zeta(2k-2t-1),
\EEQA
where we have inserted $c_0=-\frac{1}{2\beta}(4\pi)^{-3/2}$. We
notice the appearance of a temperature independent term, this
will not give any contribution to the entropy then.\\
The remaining singularity in $f_l$ for $l\geq 1$ also comes from the pole in
Riemann's zeta-function and again appears only in the $k=2$ contribution. Here 
$b_2^{(0)}$ gets multiplied by
\BEQS
    4\sqrt{\pi}c_l\beta^3h^{3/2}
    \sum_{\nu=0}^\infty(-1)^\nu\frac{z^{2\nu}}{\nu!}
    \Gamma(\nu+\frac{3}{2})\zeta(3+2\nu)(4\pi^2)^{-\nu-3/2},
\EEQS
which is non-singular, but $b_2^{(1)}$ gets multiplied by
\BEQS
    4\sqrt{\pi} c_l\frac{1}{4}
    \beta^3h^{3/2}(\zeta(1)\Gamma(\frac{3}{2})(4\pi^2)^{-3/2}+...),
\EEQS
with the non-singular terms left out. In the spirit of $\zeta$-function
regularization then, we must interpret this $\zeta(1)$ as $\gamma$, and the
result is then well behaved.\\ 
For the Minkowski space contribution we have to return to the
initial problem, the solution of the heat-equation, and we have to
redo the calculation with $l=0$. Hence
\BEQ
    f^{(0)}_0(r) = -\frac{1}{2\beta}\left.\frac{d}{ds}\right|_{s=0}
    \left(\frac{1}{\Gamma(s)}\sum_{n=-\infty}^\infty
    \int_0^\infty \sigma^{s-1}g_{n0}^{(0)}(r,\tau,r,\tau;\sigma)
    Y_{00}^2e^{-\omega_n^2\sigma} d\sigma\right),
\EEQ
where $g_{n0}^{(0)}$ solves $\Box g_{n0}^{(0)} = \omega_n^2 g_{n0}^{(0)}
-\frac{\partial}{\partial\sigma} g_{n0}^{(0)}$. So $g_{n0}^{(0)}$ is the
Minkowski space analogue of the function $g_{nl}$ we introduced in order
to solve the heat-equation in a Schwarzschild geometry. The solution
is easily seen to be
\BEQ
    g_{n0}^{(0)} = \frac{1}{\sqrt{4\pi\sigma}} e^{-\frac{(r-r')^2}{4\sigma}}
    +\mbox{terms vanishing at $r=r'$}.
\EEQ
This leads to (standard calculation)
\BEQ
    f_0^{(0)}(r) = -\frac{1}{91\pi\beta^2}.
\EEQ
We notice that this contribution is independent of $r$ and that it vanishes
as $\beta\rightarrow \infty$, it is thus a pure effect of the finite 
temperature. This also implies that there is no Minkowski spacetime 
contribution, for $l=0$,  to the pressure, but only to the entropy and 
internal energy. It is interesting to point out that in fact the entire 
Minkowski space contribution is $\beta$-dependent, also for $l\neq 0$.\\
The $l=0$ contribution to the Casimir energy density can now be written down:
\BEQA
    f_0^{\rm Cas} &=& \frac{\sqrt{h}\gamma}{16\pi^{3/2}(5-h)}(2r^{-1}-h^{-1}
      h^{'2}+h'')+\frac{1}{91\pi\beta^2}+\nonumber\\
    &&\frac{1}{8\beta\sqrt{\pi}}\sum_{k=3}^\infty\Gamma(k-\frac{1}{2})
    \left(\frac{\sqrt{h}\beta}{2\pi}\right)^{2k-1}\sum_{t=0}^{[k/2]}
    \left.b_k^{(t)}\right|_{l=0}\zeta(2k-2t-1).
\EEQA
The temperature independent part of this is negative for $r$ not too far
away from the Scwharzschild radius, and in fact looks pretty much like
the usual effective potential (i.e., Coulomb plus angular momentum part) for
the hydrogen atom, thereby suggesting the existence of bound states. We will
later see more convincing arguments for this.\\
From $f_0^{\rm Cas}$ we can evaluate the $l=0$ contribution to the 
entropy density, which turns out to be
\BEQ
    s_0 =\frac{1}{4}\pi^{-1/2}\sum_{k=3}^\infty\Gamma(k-\frac{1}{2})
    \left(\frac{\sqrt{h}\beta}{2\pi}\right)^{2k-1}\sum_{t=0}^{[k/2]}
    b_k^{(t)}\zeta(2k-t-1)(k-1-t)-\frac{2}{91\pi\beta}.
\EEQ
For $l\neq 0$ we still had a singularity, using the above prescription for its
regularization we arrive at (including the $l=0$ contribution)
\BEQA
    f_l^{\rm Cas}(r)&=&2\sqrt{\pi}c_l\left[-\frac{\gamma
    -2\ln 2\pi}{2\pi^{5/2}}\left(z^2h^{-1/2}-\tilde{z}^2-\frac{4\pi^2}{\sqrt{h}\beta^2}
    +\frac{4\pi^2}{\beta^2}\right)\right.+\nonumber\\
    &&\frac{2}{\beta\sqrt{\pi}}\sum_{\nu=2}^\infty\frac{(-1)^\nu}{\nu! (2\pi)^{2\nu-1}}
    \left(z^{2\nu}h^{-1/2}-\tilde{z}^{2\nu}\right)\Gamma(\nu-\frac{1}{2})
    \zeta(2\nu-1)+\nonumber\\
    &&\beta^3\frac{1}{2}\sqrt{\pi}\left(4b_2^{(0)}h^{3/2}\pi^{-1/2}\sum_{\nu=0}^\infty
    \frac{(-1)^\nu z^{2\nu}}{\nu!(2\pi)^{2\nu}}\Gamma(\frac{3}{2}+\nu)
    \zeta(2\nu+3)+\right.\nonumber\\
    &&b_2^{(1)}h^{3/2}(2\pi)^{-5}\left(\frac{1}{2}\gamma+\sum_{\nu=1}^\infty
    (-1)^\nu\frac{z^{2\nu}}{\nu!(2\pi)^{2\nu}}\Gamma(\nu+\frac{3}{2})\frac{1}
      {\sqrt{\pi}}
    \zeta(2\nu+1)\right)-\nonumber\\
    &&\left.\alpha_2(l)r^{-4}(2\pi)^{-3}\left\{\left(\frac{2\pi}{\tilde{z}}\right)^3
    +\frac{4}{\sqrt{\pi}}\sum_{\nu=0}^\infty\frac{(-1)^\nu\tilde{z}^{2\nu}}{\nu! (2\pi)^{2\nu}}
    \Gamma(\nu+\frac{3}{2})\zeta(2\nu+3)\right\}\right)+\nonumber\\
    &&\sum_{k=3}^\infty\Gamma(k-\frac{1}{2})\beta^{2k-1}\left(2\sum_{t=0}^{[k/2]}
    \frac{\Gamma(k-\frac{1}{2})}{\Gamma(k-\frac{1}{2}-t)}b_k^{(t)}h^{k-\frac{1}{2}}
    \sum_{\nu=0}^\infty\frac{(-1)^\nu z^{2\nu}}{\nu!(2\pi)^{2k+2\nu+2t-1}}\times\right.\nonumber\\
    &&\frac{\Gamma(k-\frac{1}{2}+\nu)}{\Gamma(k-\frac{1}{2}-t)}\zeta(2k-1+2\nu-2t)-
    \alpha_k(l)r^{-2k}(2\pi)^{1-2k}\left\{\left(\frac{\tilde{z}}{2\pi}\right)^{1-2k}
    +\right.\nonumber\\
    &&\left.\left.\left.+\frac{2}{\Gamma(k-\frac{1}{2})}\sum_{\nu=0}^\infty
    \frac{(-1)^\nu\tilde{z}^{2\nu}}
    {\nu!(2\pi)^{2\nu}}\Gamma(k+\nu-\frac{1}{2})\zeta(2k+2\nu-1)\right\}
	\right)\right]+\nonumber\\
	&& \frac{\sqrt{h}\gamma}{16\pi^{3/2}(5-h)}(2 r^{-1}-h^{-1}(h')^2 +h'')
	+\frac{1}{91\pi\beta^2}.
\EEQA
This leads to the following expression for the modes of the entropy density
\BEQA
    s_l^{\rm Cas} &=& -\beta f_l^{\rm Cas} +2\sqrt{\pi}c_l\left[
    -\frac{\gamma-2\ln 2\pi}{\pi^{5/2}}(\beta(z^2h^{-1/2}-\tilde{z}^2)+
	\frac{4\pi^2}
    {\sqrt{h}\beta}-\frac{4\pi^2}{\beta}\right.+\nonumber\\
    &&\frac{2}{\sqrt{\pi}}\sum_{\nu=2}^\infty\frac{(-1)^\nu(2\nu-1)}{\nu!(2\pi)^{2\nu-1}}
    (z^{2\nu}h^{-1/2}-\tilde{z}^{2\nu})\Gamma(\nu-\frac{1}{2})\zeta(2\nu-1)+\nonumber\\
    &&\frac{1}{2}\sqrt{\pi}\beta^4\left(4b_2^{(0)}h^{3/2}\pi^{-1/2}\sum_{\nu=0}^\infty
    \frac{(-1)^\nu z^{2\nu}(2\nu+3)}{\nu!(2\pi)^{2\nu}}\Gamma(\nu+\frac{3}{2})\zeta(2\nu+3)+
    \right.\nonumber\\
    &&b_2^{(1)}h^{3/2}\pi^{-1/2}\left(\frac{5}{2}\gamma+\pi^{-1}\sum_{\nu=0}^\infty
    \frac{(_1)^\nu z^{2\nu}(2\nu+5)}{\nu!(2\pi)^{2\nu}}\Gamma(\nu+\frac{3}{2}\zeta(2\nu+1)\right)
    -\nonumber\\
    &&\left.\frac{1}{2}\alpha_2(l)r^{-4}\pi^{-7/2}\sum_{\nu=0}^\infty
      \frac{(-1)^\nu \tilde{z}^{2\nu}(2\nu+3)}{\nu!(2\pi)^{2\nu}}
      \Gamma(\nu+\frac{3}{2})\zeta(2\nu+3)\right)+\nonumber\\
    &&\sum_{k=3}^\infty\beta^{2k}\left(2h^{k-1/2}\sum_{t=0}^{[k/2]}
      \frac{\Gamma(k-\frac{1}{2})}{\Gamma(k-\frac{1}{2}-t)}b_k^{(t)}
      \sum_{\nu=0}^\infty \frac{(-1)^\nu z^{2\nu}(2\nu+2t+2k-1)}
    {\nu!(2\pi)^{2\nu+2k+2t-1}}\times\right.\nonumber\\
    &&\frac{\Gamma(k-\frac{1}{2}+\nu)}{\Gamma(k-\frac{1}{2}-t)}
      \zeta(2\nu+2k-2t-1)-\nonumber\\
    &&2\alpha_k(l)r^{-2k}(2\pi)^{1-2k}\frac{1}{\Gamma(k-\frac{1}{2})}\sum_{\nu=0}^\infty
    \frac{(-1)^\nu\tilde{z}^{2\nu}}{\nu!(2\pi)^{2\nu}}\Gamma(k+\nu-\frac{1}{2})
	\times\nonumber\\
	&&\left.\left. (2k+2\nu-1)
    \zeta(2k+2\nu-1)\right)\right]-\frac{2}{91\pi\beta}.
\EEQA
One should note that $s_l$ is negative in some regions, this need not 
indicate that the entropy as such is negative, but only that one can not really
localise the entropy: the correct physical quantity is $S=\sum_l \int s_l r^2dr$ and not $s_l(r)$, and
this could very well be positive still. Furthermore, the entropy density is not uniquely defined as one could add
a total derivative and still get the same over all entropy. This arbitrariness
in the choice of $s_l$ will of course be fixed by defining a zero for the 
entropy proper. Furthermore, the calculation is only valid outside the 
horizon, $r\geq 2M$, so in principle one could have a negative entropy in
this part of the system (here the universe), provided that a suitable positive
entropy is present inside the horizon $r\leq 2M$. It does seem rather strange,
however, to insist on this interpretation, as $r\geq 2M$ is the only 
observable part of the universe, and we would certainly expect physical
quantities to ``behave properly'' in this region.\\
The entropy density we have found here is partly geometric in nature (induced
by the curvature) and partly thermal (coming from $\beta\neq \infty$). The
``geometric entropy'' of black holes has been studied by Moretti in 
\cite{entropy}, where he is also using $\zeta$-function techniques. He shows
that the Bekenstein-Hawking entropy is purely geometrical. We refer to his
paper for further details.\\
The $\beta$-independent part of $f_l^{\rm Cas}$ is seen to be simply
(remember, $c_l\sim \beta^{-1}, b_k^{(t)} = \beta^{-2t}c_k^{(t)}$)
\BEQA
    \left.f^{\rm Cas}_l\right|_{\rm no~~temp.}&=& 2\sqrt{\pi}\beta c_l 
	\frac{1}{2}\sqrt{\pi} c_2^{(1)} h^{3/2} (2\pi)^{-5} \frac{1}{2}
	\gamma+\frac{\sqrt{h}\gamma}{16\pi^{3/2}(5-h)}(2r^{-1}-h^{-1}(h')^2
	+h'')\nonumber\\
	&=& \frac{(2l+1)}{2^l 64 \pi^{7/2}} M^2r^{-2} (r-2M)^{-1/2} (2r+M)^{-1}
	\sum_{\nu=0}^{[l/2]}\frac{(-1)^\nu (2l-2\nu)!}{\nu !(l-\nu)!(l-2\nu)!}
	+\nonumber\\
	&&\frac{1}{16\pi^{3/2}}\sqrt{h}\gamma (5-h)^{-1} (2r^{-1}-h^{-1}(h')^2+
	h''),
\EEQA
(the last term coming from $l=0$)
which is divergent on the horizon. By using
\BEQS
	1=\sum_{lm} |Y_{lm}|^2 = \frac{1}{4\pi}\sum_{l=0}^\infty\frac{2l+1}
	{2^l}\sum_{\nu=0}^{[l/2]} \frac{(-1)^\nu (2l-2\nu)!}{\nu!(l-\nu)!
	(l-2\nu)!},
\EEQS
which follow from the formula for $P_l(1)$ used before, we can actually carry
out the sum over $l$ to obtain
\BEQ
	\left.f^{\rm Cas}\right|_{\rm no~temp} = \frac{M^2\gamma}{16\pi^{5/2}}
	r^{-2}(r-2M)^{-1/2} (2r+M)^{-1} +\frac{\sqrt{h}\gamma}{16\pi^{3/2}
	(5-h)}(2r^{-1}-h^{-1}(h')^2+h''). \label{eq:fnotemp}
\EEQ
It is also interesting to note that outside the Schwarzschild radius this is
positive, whereas as inside it is negative imaginary. The integral, moreover,
over $r$ is finite (for $l\neq 0$) in any case, even though the function 
is divergent for $r=2M$, and we get the following result
\BEQ
	\left.F\right|_{\rm no~~ temp} = \frac{\gamma}{16\pi^{5/2}}\left[
	\frac{\pi}{\sqrt{10 M}} -i\sqrt{\frac{2}{5M}}
	{\rm Artanh}\left(\frac{2}{\sqrt{5}}\right)\right] \qquad l\geq 1,
\EEQ
the first term is the $r> 2M$ contribution, whereas the second comes from
$0<r<2M$.
This result represents, then the energy coming from the removal of the
point $r=0$ from Minkowski space and placing a mass $M$ in that singularity. It
is, in other words, a proper Casimir energy similarly to the one one obtains
in the original case of two plates in flat spacetime.\\
From (\ref{eq:fnotemp}) we can also find a temperature independent pressure
by simply taking the derivative with respect to $r$. Doing this we find
\BEQA
	\left. p\right|_{\rm no~temp} &=&- 
	\frac{M^2\gamma}{64\pi^{7/2}}
	\frac{8 M^2+19 Mr-14 r^2}{2 r^4 h^{3/2}(2r+M)^2}-\nonumber\\
	&&\frac{\gamma}{64\pi^{5/2}}\frac{10M^4+16 M^3 r-32 M^2 r^2-2 M^3 r^2
	+12Mr^3-11M^2r^3+10Mr^4-2r^5}{r^5 h^{3/2} (2r+M)^2}
\EEQA
This local pressure is negative outside the horizon, i.e., for  $r>2M$, 
and negative imaginary for $r<2M$. \\
For small but non-vanishing $\beta$ we get (writing $c_l=\beta^{-1}
\bar{c}_l, z^2=\beta^2Z^2, \tilde{z}^2=\beta^2\tilde{Z}^2$)
\BEQA
	f_l^{\rm Cas} &=& 2\sqrt{\pi}\bar{c}_l \left[-\frac{\gamma-2\ln 2\pi}
	{2\pi^{5/2}}\beta(Z^2 h^{-1/2} -\tilde{Z}^2-\beta^{-4}4\pi^2
	(h^{-1/2}-1))+\right.\nonumber\\
	&&\frac{1}{8}\pi^{-7/2}\beta^2 (Z^2 h^{-1/2}-\tilde{Z}^2)\Gamma(3/2)
	\zeta(3)+\frac{1}{2\pi}c_2^{(0)}h^{3/2}\beta^2\Gamma(3/2)\zeta(3)+
	\nonumber\\
	&& (2\pi)^{-5} h^{3/2} c_2^{(1)}\left(\frac{1}{2}\gamma-(2\pi)^{-2}
	Z^2\beta^2\Gamma(5/2)\zeta(3)\right)-\nonumber\\
	&&\alpha_2(l) r^{-4}(2\pi)^{-3}\left((2\pi)^3\tilde{Z}^{-3}\beta^{-4}
	+\frac{4}{\sqrt{\pi}}\beta^2\Gamma(3/2)\zeta(3)\right)-\nonumber\\
	&&\left.\sum_{k=3}^\infty\Gamma(k-\frac{1}{2})\alpha_k(l) r^{-2k} 
	\tilde{Z}^{1-2k}\beta^{-1}\right] +\nonumber\\
	&&\frac{\sqrt{h}\gamma}{16\pi^{3/2}(5-h)}(2r^{-1}-h^{-1}(h')^2+h'')
	+\frac{1}{91\pi\beta^2}
	+O(\beta^3).
\EEQA
Note that in the summation over $k$ contribution, the summation does not 
extend to a summation over powers of $\beta$; $\beta^{-1}$ is multiplied by a
function of $r,l$, $A_l(r) := \sum_{k=3}^\infty \Gamma(k-1/2)\alpha_k(l)
r^{-2k}\tilde{Z}^{1-2k}$. The free energy resulting from this expression
turns out to be everywhere negative for $r>2M$ and imaginary for
$r<2M$.\\
The entropy density we derive from this is
\BEQA
	s_l^{\rm Cas} &=& 2\sqrt{\pi}\bar{c}_l\left[-\beta^2
	\frac{\gamma-2\ln 2\pi}
	{2\pi^{5/2}}(Z^2h^{-1/2}-\tilde{Z}^2)-6\beta^{-2}\frac{\gamma-
	2\ln 2\pi}{\sqrt{\pi}}(h^{-1/2}-1)\right.+\nonumber\\
	&&\frac{1}{8}\pi^{-3}\zeta(3)\beta^3 (Z^2h^{-1/2}-\tilde{Z}^2)
	+\frac{1}{2}\pi^{-1/2}c_2^{(0)}h^{3/2}\zeta(3)\beta^3-\nonumber\\
	&&\left.\frac{3}{2}\sqrt{\pi}(2\pi)^{-7}Z^2h^{3/2}c_2{(1)}\zeta(3)
	\beta^3
	+4\alpha_2(l)r^{-4}(\tilde{Z}^{-3}\beta^{-3}-
	\zeta(3)(2\pi)^{-3}\beta^3)+A_l(r)\right]-\nonumber\\
	&&\frac{2}{91\pi\beta}+O(\beta^3).
\EEQA
This is positive outside the Schwarzschild radius.
We see that the temperature independent part of the entropy density is
simply $A_l(r)$ for $l\geq 1$ which is independent of the mass of the black 
hole and comes from the renormalisation, and the $l=0$ contribution then 
has all the information about $M$. As well the free energy as the entropy
and internal energy densities are divergent on the horizon $r=2M$,
although in a rather mild way.\\
The internal energy density is found to be
\BEQA
	u_l^{\rm Cas} &=& 2\sqrt{\pi}\bar{c}_l\left[-(\gamma-2\ln 2\pi)
	\pi^{-5/2}\beta^3(Z^2h^{-1/2}-\tilde{Z}^2)-4\pi^{-1/2}(\gamma-2
	\ln 2\pi)\beta^{-1}(h^{-1/2}-1)+\right.\nonumber\\
	&& \frac{3}{16\pi^3}\beta^4 \zeta(3)(Z^2h^{-1/2}-\tilde{Z}^2)
	+\frac{3}{4\sqrt{\pi}}c_2^{(0)}\zeta(3)h^{3/2}\beta^4+\nonumber\\
	&&(2\pi)^{-5}h^{3/2}c_2^{(1)}(\frac{1}{2}\gamma\beta^2+\frac{9}{16
	\pi^{3/2}}Z^2\zeta(3)\beta^4)-\nonumber\\
	&&\left.\alpha_2(l)r^{-4}(-3\tilde{Z}^{-3}\beta^{-2}+
	\frac{4}{3\pi^3}\zeta(3)\beta^4)\right]+\nonumber\\
	&&\frac{\sqrt{h}\gamma}{16\pi^{3/2} (5-h)}(2r^{-1}-h^{-1}(h')^2
	+h'')\beta^2-\frac{1}{91\pi}+O(\beta^5),
\EEQA
which is seen to be singular at $r=2M$ as mentioned above. It is interesting
to note that, contrary to the free energy and the entropy densities, the 
internal energy is integrable inside the horizon. In fact we get
\BEQA
	\left.U_l^{\rm Cas}\right|_{\rm int} &=& \int_0^{2M}dr r^2u_l^{\rm Cas}
	\\
	&=&-12 M^2 (l(l+1))^{-1/2}\beta^{-2}+\frac{\gamma M\ln 5}{128\sqrt{5}
	\pi^3}(24 \pi^{3/2}-32-5M\pi^{3/2})\beta^2+\nonumber\\
	&&\frac{8}{273\pi}M^3\beta^{-1}(364\gamma\sqrt{\pi}-\beta-728\sqrt{\pi}
	\ln 2\pi)-\nonumber\\
	&&i\frac{M}{128\sqrt{\pi}}\beta^{-1}
	(1280 M^2\pi(\gamma-2\ln 2\pi)+\gamma\beta^3 (M-8))-\nonumber\\
	&&(i\pi+\ln 5)\frac{\gamma M}{128\sqrt{5}\pi^3}\beta^2(32-24\pi^{3/2}
	+5M\pi^{3/2}),
\EEQA
which is clearly complex but finite. The imaginary part of this internal
energy is interpreted as giving a continous creation of particles. Notice, by
the way, that only one of the terms in $U_l^{\rm Cas}$ is $l$-dependent.
The temperature independent part, moreover, is seen to be $-\frac{8M^3}
{273\pi}$, a negative constant. In any case, for all values of $\beta$ there
is a region in which $U_l^{\rm Cas}<0$, the size of this $r$-interval,
however, becomes rapidly smaller as $\beta$ increases. The actual depth of the
resulting potential ``well'', moreover, also rapidly decreases as $\beta$
grows. But for $\beta\approx .1$ a very large range of $r$ values exist
for which the internal energy is negative, once more pointing towards the
possible existence of bound states. This region only exist for $l<10$, though.
For higher values of angular momentum, no such region seems to exist, 
suggesting that only orbits with low angular momentum can be bound, which is
what one would expect on physical grounds anyway.\\
Inside the Schwarzschild radius both the
real and imaginary parts of the internal energy density seem to be 
negative for all values of $\beta$, with the energy density becoming almost
entirely a large negative imaginary number near the horizon and a large
negative real number near the origin. This apparently holds for all values of
$l$.

\section{Interpretation of $f^{\rm Cas}$}
We can find a closed expression for the summations over powers of $z$, 
by noting with Ramond \cite{Ram}
that
\BEQS
    \sigma^{-d/2} = \frac{2^{\frac{d+1}{2}}}{\sqrt{\pi}(d-2)!!}\int_0^\infty 
    e^{-\sigma w^2}w^{d-1}dw.
\EEQS
Inserting this at an earlier stage (i.e. before the $\sigma$-integration) we 
arrive at the following closed formula
\BEQ
    \sum_{\nu=0}^\infty\frac{(-1)^\nu y^{2\nu}}{\nu!}\Gamma(\nu-\frac{1}{2})
    \zeta(2\nu-1) \propto -\beta^{-1}\pi^{-3/2}\int_0^\infty \sqrt{w^2+y^2}
    +2\ln\left(1-e^{-\sqrt{w^2+y^2}}\right)dw,
\EEQ
where
\BEQ
    y^2=\frac{z^2}{2\pi} = \frac{l(l+1)h(r)\beta^2}{2\pi r^2}.
\EEQ
Now, this is essentially the expression for the free energy of a scalar
quantum field in $d=1$ dimension (see e.g. \cite{Ram}) and with the energy of 
the 
individual quanta being given by $\omega^2=y^2/\beta^2=l(l+1)h(r)/(2\pi r^2)$.
Our Casimir energy density is hence
the regularised expression for the energy of an infinite family (labelled by
their angular momentum quantum number $l$) of such quanta.\\
This was to be expected, since the spherical symmetry of the system implies
that the problem is essentialy unidimensional (the radial coordinate). 
Moreover, the angular quantum number $l$ appears as defining an $r$-dependent
mass.\\
Expanding to first order in $M$ we then arrive at expressions of the form
(valid only for $l\neq 0$)
\BEQA
    f^{\rm Cas}_l&=&\beta^{-1}\pi^{-3/2}\int_0^\infty\left[
      \sqrt{w^2+\frac{l(l+1)\beta^2}{2\pi r^2}}
    +2\ln\left(1-e^{\sqrt{w^2+\frac{l(l+1)\beta^2}{2\pi r^2}}}\right)
	\right]dw\nonumber\\
    &&-M\frac{\beta l(l+1)}{2\pi^{5/2}r^4}\int_0^\infty 
      (w^2r^2+\beta^2 l(l+1)))^{-1/2}\times\nonumber\\
    &&\left(1+2
    \frac{e^{-\sqrt{z^2+l(l+1)\beta^2/r^2}}}{1-
      e^{-\sqrt{w^2+l(l+1)\beta^2/r^2}}}
    \left(w^2+\frac{l(l+1)\beta^2}{2\pi r^2}\right)^{-1/2}\right)dw+O(M^2)
	.\nonumber\\
\EEQA
The first two terms are the contribution coming from a massless particle 
in a flat one dimensional space. While the second term shows how the 
pressence of a gravitational field (here $M$) modifies the energy. 

\section{Discussion and Conclusion}
We have obtained asymptotic expressions for the heat kernel and thereby 
for the free 
energy density in a Schwarzschild geometry by use of the $\zeta$-function 
technique. We
also took the $M\rightarrow 0$ limit, i.e. the flat space limit, and we
subtracted the two energies, to obtain what we called the Casimir energy
density, this is the part of the zero point energy density due to the 
curvature (i.e. to the deviation from Minkowski spacetime) and
is thus an intrinsically interesting quantity. As we would expect, these
densities all turned out to depend only on the radial coordinate.\\
The major unanswered question, however, concerns the boundary conditions.
The asymptotic expansion of the heat kernel is independent of the chosen
boundary conditions, i.e. on the particular vacuum state. It is known, however,
that the full renormalised energy-momentum tensor is sensitive to the 
particular vacuum state. Page, \cite{Page}, has computed the energy momentum
tensor in a Schwarzschild background for conformal coupling, and he got
the $00$ contribution to be
\BEQAS
	T^0_0 &\approx& (-9216 M^{10}-21504 M^9 r- 18688 M^8 r^2+7168 M^7
	r^3 -\\
	&&2560 M^6 r^4 + 4 M^2 r^8 - 4M r^9 + r^{10})/(122880 M^2
	r^{10} \pi^2)
\EEQAS
Anderson et al, \cite{AHWY} have extended the original calculation by Page to
arbitrary coupling to curvature in the Hartle-Hawking vacuum state, and they
then find a finite energy density at the horizon $\rho(r=2M)=(15\xi-4)/(15360
M^4\pi^2)$. They also find that $\rho$ is positive for $r>2M$ only
for $\frac{4}{15}<\xi <1.2575$, where $\xi$ is the non-minimal coupling.
In our case we have $\xi=0$, and we found the Helmholtz free energy to be
positive for $r>2M$ in the limit of vanishing temperature. We also found, 
however, the internal energy to be divergent at the Schwarzschild radius,
and, moreover, to posses regions in which it was negative, suggesting the
existence of bound states.\\
Our computation seems to have more in common with Boulware vacuum. Candelas, 
\cite{Candelas}, has computed the renormalised value of $\langle\phi^2\rangle$
in the Boulware, Unruh and Hartle-Hawking vacuum states, and found that for
$r\rightarrow 2 M$ the Boulware vacuum expression diverges, whereas the others
are finite. Furthermore, as $r\rightarrow \infty$, $\langle\phi^2\rangle$ 
vanishes in the Boulware vacuum and not in the otehr two. Both of these
features are reproduced by the present calculation, which therefore seems
to be closer related to the Boulware vacuum than any of the other two
known vacuum states. It is impossible to tell precisely, however, since
the Schwinger-DeWitt expansion which we used partially in obtaining our
result is insensitive to the details of the vacuum state. This does not imply,
though, that the present calculation is meaningless from a physical point
of view, since Anderson and coworkers, \cite{A2}, have shown that for
$(r-2M)/M$ not too large, the Schwinger-DeWitt expansion expression for
$\langle T_{00}\rangle$ is valid, and only for $(r-2M)/M\gg 1$ (i.e., $r\gg 
3M$) does the
state-dependence become noticable. Hence, for $(r-2M)/M \lesssim 1$ at 
least,
the present calculation is valid. Moreover, since we use a hybrid technique
mixing the exact mode sum with the asymptotic Schwinger-DeWitt expansion,
one can expect the region of validity of this calculation to be larger than
the pure Schwinger-Dewitt approach. It remains for future research to 
specify the precise range of validity, to see just how far away from $r
\approx 3M$ we can push the present computation.\\
All the
calculations were carried out at finite temperature. The quantities we have
found are therefore also the effective actions for a free (minimally coupled)
scalar field in these backgrounds. The Casimir density then shows how the
presence of curvature modifies the effective action for a flat spacetime.\\
We noticed, for instance, the presence of a constant (with respect to $r$), 
positive contribution to the entropy, which we can interpret as generation of 
radiation, somewhat related to the Hawking radiation but not quite the same, 
as this time the radiation clearly came from the distortion of the surrounding 
vacuum and could not be attributed to the internal (quantum) structure of the
massive object generating the Schwarzschild geometry. This contribution,
moreover, depended upon the temperature (was, in fact, proportional to it)
and can in this way be seen as a correction to the Hawking radiation, by
letting the temperature be equal to the Hawking temperature $T_H$.\\
Especially interesting would be the coupling of the Casimir energy density to
the curvature, i.e. plugging in the Casimir energy density into the right hand
side of Einstein's field equations and assume $M=M(r,t)$ is a sufficiently
slowly varying function of $r,t$ (if not, one will have to redo the entire
calculation of $f^{\rm Cas}$ with a time and radial dependent mass). Will the
singularity at $r=0$, stemming from the point mass ($M(r,t)=M\delta(r)$
independent of $t$), become ``dressed'' in this way? If so, then quantum
effects could be responsible for the removal of other singularities as well,
most notably the singularity at the Big Bang (or the Big Crunch). Will this
constitute a general ``quantum removal of singularities''-mechanism? This back
reaction would also give us a more precise picture of the internal structure of
a black hole, as well as of its stability; we already know that black holes can
evaporate due to Hawking radiation, but the calculations put forward here
suggests the existence of even more such effects, this time coming directly
from the disturbance of the surrounding vacuum, from the ``dressing'' of the
black hole so to speak. This means that at least part of the Bekenstein-Hawking
entropy must be of a geometrical nature, consequently supporting Moretti's
findings, \cite{entropy}. The fact that we get negative values for the 
candidate entropy density shows that this entropy cannot be localised in one
particular region (such as near the Schwarzschild radius, say) but has to be
considered a global object intimately related to the global topological
properties of the spacetime manifold.

\appendix
\section{The Explicit Form of the $b_k^{(t)}$}
In this appendix we list, for the readers convenience, the explicit form
of the first few $b_k^{(t)}$. For $k=0$ we have simply $b_0^{(0)}=1$ while
for $k=1$ we have $b_1^{(0)}=0$. All of these coefficients can be written
as rational functions of $r,l,\beta$ and $M$.

\subsubsection{The Case of $k=2$}
Here we have two possible values of $t$, namely $t=0,1$ and the
functions are readily found to be
\BEQAS
    b_2^{(0)} &=& -2l(l+1)\frac{r-4M}{2r^5+Mr^4}\\
    b_2^{(1)} &=& -\frac{16\pi^2 M^2}{\beta^2 r(r-2M)^2(2r+M)}
\EEQAS
We note the appearance of a zero at $r=4M$ in $b_2^{(0)}$ and a pole of order
two at $r=2M$ in $b_2^{(1)}$
 
\subsubsection{The Case of $k=3$}
Again we have only two coefficients, corresponding once more to $t=0,1$.
The functional expressions, however, become somewhat more complicated.
\BEQAS
    b_3^{(0)} &=& l(l+1)\frac{96r^4-640 Mr^3+156 M^2r^2+288 M^3r+64M^4}
    {r^6(2r+M)^3(3r+M)}\\
    b_3^{(1)} &=& \pi^2M^2\frac{768 r^4+512Mr^3+244M^2r^2-576M^3r-128M^4}
    {\beta^2r^3(r-2M)^3(2r+M)^3(3r+M)}
\EEQAS
Here there are no zeroes but only a pole of order three at $r=2M$.

\subsubsection{The Case of $k=4$}
For $k=4$ we get very complicated expressions for all three possible
values of $t$, $t=0,1,2$.\\
The numerator for $t=0$ is
\BEQAS
    &&l(l+1)\{(8640-103680l(l+1))r^9+(995328-37152l(l+1))r^8M-\\
    &&(532800+53712l(l+1))r^7M^2-(1407552-79016l(l+1))r^6M^3+\\
    &&(71328+207464l(l+1))r^5M^4+(1035336+182480l(l+1))r^4M^5+\\
    &&(717648+85024l(l+1))r^3M^6+(228192+22504l(l+1))r^2M^7+\\
    &&(36096+3208l(l+1))rM^8+(2304+192l(l+1))M^9\}
\EEQAS
while the denominator is
\BEQAS
    &&r^8\{M^9+23M^8r+233M^7r^2+1365M^6r^3+5098M^5r^4+\\
    &&12592M^4r^5+20576M^3r^6+21456M^2r^7+12960Mr^8+3456r^9\}
\EEQAS
For $t=1$ the numerator becomes
\BEQAS
    &&\pi^2M\{768l(l+1)M^{11}+(12032l(l+1)+1536)M^{10}r
    +(76160l(l+1)+24064)M^9r^2+\\
    &&(240384l(l+1)+150272)M^8r^3+(350256l(l+1)+142976)M^7r^4+\\
    &&(37968l(l+1)-717248)M^6r^5-(469392l(l+1)+436992)M^5r^6-\\
    &&(265040l(l+1)-1506816)M^4r^7+(310048l(l+1)-993792)M^3r^8+\\
    &&(124992l(l+1)-884736)M^2r^9-(141696l(l+1)+829440)Mr^{10}
    +27648r^{11}\}
\EEQAS
whereas the denominator reads
\BEQAS
    &&\beta^2r^5\{16M^{13}+336M^{12}r+3016M^{11}r^2
    +14928M^{10}r^3+\\
    &&43297M^9r^4+69255M^8r^5+37937M^7r^6-52347M^6r^7-81046M^5r^8+\\
    &&3504M^4r^9+49376M^3r^{10}+720M^2r^{11}-14688Mr^{12}+3456r^{13}\}
\EEQAS
For the last case, $t=2$, the numerator is simply
\BEQS
    \pi^4M^3(1024r-128M)
\EEQS
while the denominator, a bit more complicated, reads
\BEQS
    \beta^4r(r-2M)^4(2r+M)^2(4r+M)
\EEQS

\end{document}